\documentclass[twocolumn,prb,superscriptaddress,aps,10pt]{revtex4-1} 

\usepackage{amsmath}  
\usepackage{amsfonts} 
\usepackage{graphicx} 
\usepackage{hyperref}
\usepackage{enumitem}
\usepackage{minibox}
\usepackage{color}

\begin{document}


\title{A Model for Incorporating Computation Without Changing the Course:\\An example from middle-division classical mechanics}

\author{Marcos D. Caballero}
\email{Corresponding author: caballero@pa.msu.edu} 
\affiliation{Department of Physics and Astronomy, Michigan State University, East Lansing, MI 48824, USA} 

\author{Steven J. Pollock}
\affiliation{Department of Physics, University of Colorado Boulder, Boulder, CO 80302-0390}


\date{\today}

\begin{abstract}
Much of the research done by modern physicists would be impossible without the use of computation.
And yet, while computation is a crucial tool of practicing physicists, physics curricula do not generally reflect its importance and utility. 
To more tightly connect undergraduate preparation with professional practice, we integrated computational instruction into middle-division classical mechanics at the University of Colorado Boulder.
Our model for integration works within the constraints of faculty who do not specialize in computation teaching standard physics courses by placing a strong emphasis on {an adaptable curriculum}. 
Our model includes the construction of computational learning goals, the design of computational activities consistent with those goals, and the assessment of students' computational fluency.
We present critiques of our model as we work to develop an effective and sustainable model for computational instruction in the undergraduate curriculum.

\end{abstract}

\maketitle 

\section{Introduction}\label{sec:intro}

Modern scientists increasingly rely on the predictions of computational models to explore and to understand the natural world.
Scientists use computation to model complex physical systems and to gain insight about the systems' properties (e.g., chaotic dynamics).\cite{DasBuch}
Experiments that are impossible to perform can be simulated (e.g., star formation)\cite{kim2009galaxy} while data that are too mountainous to imagine are reduced to a sensible size (e.g., LHC data).\cite{duhrssen2004extracting}
Given its ubiquitous use in science, computational modeling has been described as the ``third leg'' of modern science.\cite{denning2007computing}
Physics educators recognize the importance of teaching computational modeling; but, the courses taught by these educators ignore or under-emphasize this crucial tool.\cite{Chonacky:2008gq}
Hence, many physics bachelor's graduates are not prepared to use computational modeling in their future careers.\cite{Ivie:2002pb}



In undergraduate physics, computational modeling is often relegated to a single upper-division elective course, which may be taken asynchronously with students' experience -- prior to learning important physics or mathematical content needed in this course or learning computation well after it could have been useful.
Furthermore, these computational physics courses usually devote significant time to underlying algorithms, error analysis, and optimization while sacrificing discussions of when and why to use computational modeling.
A deep understanding of computational modeling is important for students who choose to become computational specialists, but for other students, simply gaining a familiarity with a computational environment and knowing when and why to use it is more important.

Recently, the American Journal of Physics published a special two-issue edition in which a number of different courses that incorporated computational modeling (in varying ways) were discussed. \cite{AJPcompIssue}
Most articles detailed the computational environments used and the problems that students encountered. 
However, the success of these and other implementations has relied chiefly on singular champions,\cite{timberlake2008computation} has exploited small class sizes,\cite{roos2006incremental} or has required drastic, under-documented changes to the instructional environment that have not been adopted widely.\cite{mcintyre2008integrating}
Lacking from these encouraging changes is an instructional model that employs a strategic approach that non-computational specialist faculty can easily adopt and adapt for their own courses without significant overall changes to the course structure and class size.
A blueprint for integrating computational modeling into lecture courses, not simply a list of topics or problems, is needed to develop and broaden computational instruction in the undergraduate curriculum.
Furthermore, while explicating an instructional model is important to start a reform, student buy-in is critical for sustaining it.\cite{dancy2013}
Students' perspectives on these courses and on computational modeling more generally have gone underreported in previous literature.

This paper addresses the shortcomings of previous work by describing one model for computational instruction that can be incorporated into standard physics courses and used by non-specialist faculty. We provide details about the learning goals that guided the construction of instructional materials and the design of an open-ended computational assessment. Students' reflections on learning computational modeling in this course are also reported. We close with our own reflections as we look to develop a sustainable model for computational instruction in undergraduate physics courses.

\section{Motivation \& Philosophy}\label{sec:motiv}

We integrated computational modeling into a middle-division classical mechanics course at the University of Colorado Boulder (CU Boulder).
Computational modeling in this courses helped to extend coursework beyond the analytically tractable, to provide helpful visual elements that can facilitate deeper understanding, and to incorporate aspects of authentic, scientific work \cite{Chabay:2008jw} -- that is, it helped to enculturate students using the work of practicing physicists.
Our weaving computation into the course was catalyzed by our students' recognition of its importance in their future work; 95\% of our physics majors recognized computational modeling was at least ``somewhat important'' to their future.\cite{cucomp}
Moreover, our physics major already requires the maximum amount of coursework that the University will allow.

An instructional model that incorporated some aspects of computational modeling, but still left the course nearly intact was key for developing a model that non-specialist faculty would adopt.
The instructional time is finite and, thus, difficult choices were made.
For example, teaching students to use computational error analysis to analyze their simulations requires substantial effort and emphasis.\cite{roos2006incremental,timberlake2008computation}
Thus, we supplanted  error analysis with checks such as limiting cases to help students gain confidence that their computational model was working properly.
{Adopters might decide that the aspects we avoided teaching are important for their courses. Our instructional model (Secs.\ \ref{sec:goals}-\ref{sec:proj}) is flexible and can be easily adapted to include these aspects.} 
A single upper-division computational physics course still makes sense as a capstone course that provides students with a deeper understanding of the tools and algorithms they have learned along the way.

\section{Instructional Setting}\label{sec:is}

\begin{figure}[t]
\centering
\includegraphics[width=0.85\columnwidth]{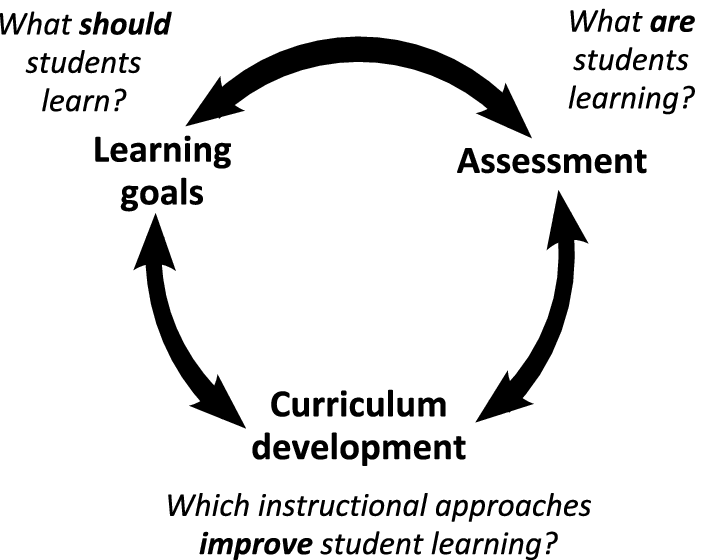}
\caption{The three coordinating aspects of the CU Science Education Initiative course transformation model.\cite{Chasteen:2009wn} This model was leveraged to construct computational learning goals (Sec.\ \ref{sec:goals}), to develop instructional materials and methods (Sec.\ \ref{sec:act}), and to assess student outcomes (Sec.\ \ref{sec:proj}).}\label{fig:sei}
\end{figure}

Classical Mechanics and Mathematical Methods 1 (CM 1) at CU Boulder is the first half of a two-semester sequence in classical mechanics usually taken in the second semester of students' sophomore year.
CM 1 is a gateway course; it is the first courses in the major sequence taken primarily by students majoring or minoring in physics, engineering physics, or astrophysics.
Prior to this course, students have taken three general physics courses: introductory calculus-based mechanics and electromagnetism, as well as modern physics.
Instruction in CM 1 is student-centered and includes electronic voting systems (``clicker questions'')\cite{banks:2006au} often with Peer Instruction,\cite{mazurpeer,crouch2001peer} tutorial-style activities,\cite{tutorialswash, ambrose:2013im} and redesigned homework problems that emphasize reasoning.\cite{hammer1996more}
Pedagogical changes have followed the Science Education Initiative's (SEI) evidence-based transformation model (Fig.\ \ref{fig:sei}).\cite{Chasteen:2009wn,Pollock:2012uy,2012AIPC.1413..291P} 


In spring 2012, CM 1 was offered to 70 students, which comprise the sample for all data presented in Secs.\ \ref{sec:act}--\ref{sec:refs}. 
The course met twice per week for 75 minutes in a typical auditorium-style classroom. 
Almost all students taking CM 1 had some prior exposure to computational modeling in a previously taken, applied math course in which the instructor assigned basic plotting tasks in either Mathematica or MATLAB.
However, by their own admission, most students received minimal instruction and were uncomfortable with using computation as we intended (e.g., solving differential equations numerically). 
Computational instruction during CM 1 lecture was minimal (Sec.\ \ref{sec:act}) because we adhered to the common course schedule that had been established two years prior.

\section{Computational Learning Goals}\label{sec:goals}

\begin{figure*}

\fbox{
\parbox{6.5in}{
{\bf Course-scale goals -- Students should be able to:}
\begin{itemize}[noitemsep,nolistsep]
	\item Map physical situations onto computational tasks.
	\item Identify the appropriate computational tools for answering a particular question.
	\item Implement computational methods to facilitate an answer.
	\item Reflect on the computational output to determine if a satisfactory answer has been constructed.
\end{itemize}
}}

\fbox{
\parbox{6.5in}{
{\bf Topic-scale goals - Students should be able to:}
\begin{itemize}[noitemsep,nolistsep]
	
	\item use the computational environment as a calculator
	
	\begin{itemize}[noitemsep,nolistsep]

		\item represent numbers, complex numbers, and vectors in the computational environment
		\item perform appropriate mathematical operations on these numbers and vectors
	
	\end{itemize}

	\item assign and use variables; develop and implement simple functions

	\begin{itemize}[noitemsep,nolistsep]

		\item access constants and physical constants defined by the computational environment
		\item define constants and variables and access those values as needed
		\item write and use simple functions

	\end{itemize}

	\item perform algebraic manipulations and solve algebraic equations

	\begin{itemize}[noitemsep,nolistsep]

		\item find the roots of an algebraic equation using a root finder by setting appropriate neighborhoods
		\item solve for an unknown value in an algebraic equation using a “solve” function by setting appropriate domains
		\item use a trigonometric reduction or simplification function appropriately

	\end{itemize}

	\item plot quantities and functions

	\begin{itemize}[noitemsep,nolistsep]

		\item plot data and functions of one or two variables
		\item plot solutions to ODEs and PDEs
		\item plot vector fields and contour lines

	\end{itemize}

	\item use the computational environment to solve differential equations

	\begin{itemize}[noitemsep,nolistsep]

		\item integrate a complicated expression (symbolically) 
		\item integrate a complicated expression (numerically)
		\item solve an ordinary differential equation with given initial conditions (symbolically)
		\item solve an ordinary differential equation with given initial conditions (numerically)
		\item solve a set of coupled ordinary differential equations with given initial conditions (numerically)

	\end{itemize}

\end{itemize}
}}
\caption{Computational learning goals developed for CM 1. Course-scale goals are broadly applicable to the practice of computational modeling, while topic-scale goals focus on basic fluency with the computation environment.}\label{fig:goals}
\end{figure*}


To begin integrating computational modeling into CM 1, we implemented the SEI course transformation model (Fig.\ \ref{fig:sei}) by first constructing computational learning goals.
Because we wanted these learning goals to be useful beyond our department, we invited faculty from other institutions to help construct and critique our goals.
These faculty helped us navigate the undesirable situation of emphasizing details of our chosen computational environment (Mathematica) over the practice of computational modeling.
We chose to use Mathematica because of its availability to our students, its familiarity to our faculty, and its perceived low-learning curve. 

Our computational learning goals exist at two scales.
{\it Course-scale goals} are widely applicable to the practice of computational modeling, for example, ``{\it students should be able to map physical situations to computational tasks.}''
{\it Topic-scale goals} focus on particular practices within physics courses, for example, ``{\it students should be able to numerically solve ordinary differential equations with given initial conditions.}''
Given that most students taking CM 1 had little instruction in computation, the topic-scale goals we constructed aim to develop students' basic fluency with computational modeling.
These goals are platform-agnostic and can be used regardless of the chosen computational environment (e.g., Mathematica, MATLAB, Maxima, Python).
The complete list of topic-scale goals we developed is applicable to a number of physics courses, but only those goals that were used in CM 1 are presented (Fig.\ \ref{fig:goals})


\section{Computational Instruction: Activities and Practices}\label{sec:act}

\begin{figure*}[t]

\fbox{
\parbox{6.5in}{\flushleft On Mythbusters, Jamie and Adam considered if a penny dropped from the Empire State building (381 m) can kill a person walking below (Episode clip: \url{http://youtu.be/PHxvMLoKRWg}).  Let's do our own investigation! 

\begin{enumerate}[noitemsep,label=(\roman*)]
\item Find the terminal speed of a dropped penny, taking into account both linear and quadratic terms. 
Write down the differential equation for the motion of a falling penny. Keep both the linear and quadratic terms. 

\item For what range of speeds will (i) the linear term dominate? (ii) the quadratic term dominate?  
If you had to pick just one term, which would you use, and why? 

\item Keeping both linear and quadratic terms gives a non-linear differential equation, which we can only solve numerically.  Use a numerical differential equation solver to determine and plot velocity and position of the penny as it falls.  
Find the time it hits the ground, and the speed with which it hits. 
Include your code with your homework.
Mathematica users might find this screencast helpful:
\url{http://youtu.be/zKO6v0w0KdI}

\item Compare your numerical result for ``final velocity" (part iii) with what your answer to part i, and also to what you get by assuming the one dominant drag term you chose in part ii. Comment. 

\item  So, could this penny kill someone? Use freshman physics to make crude (but quantitative) estimates to discuss whether you think the myth is busted or not. (There's about 1\:mm of tissue covering the skull.)
 
\end{enumerate}}}
\caption{Abbreviated version of a homework problem given in the 3$^{\mathrm{rd}}$ week of class while students are learning about models of drag and how to solve ordinary differential equations. All homework questions as they appeared to students appear online.\cite{cucomp}}\label{fig:hw3}
\end{figure*}

Prior work in computational instruction has exploited small class sizes,\cite{roos2006incremental,timberlake2008computation} has fundamentally reconfigured the curriculum,\cite{mcintyre2008integrating} or has utilized laboratory sections staffed by well-trained teaching assistants.\cite{beichner2010labs,Caballero:2011tk,Caballero:2012hu}
With higher student-to-instructor ratios, no laboratory or recitation section, and a need to cover the same sophisticated (analytical) content, we developed an instructional model that leveraged aspects of the ``flipped classroom.''
In a flipped classroom, lectures are recorded and viewed by students prior to coming to class. 
Class time is then spent discussing students' difficulties and working problems in small groups.\cite{Lage:2000fl} 
In our course, we recorded screencasts of MDC working through a variety of problems.\cite{cucomp}
Less than one hour of planning, recording, and editing was required to produce a single, polished five minute screencast.
Seven screencasts provided instruction on particular computational tools such as root finders (\texttt{FindRoot}) and numerical differential equation solvers (\texttt{NDSolve}).

Consistent with the SEI model (Fig.\ \ref{fig:sei}),\cite{Chasteen:2009wn} homework problems were developed from computational learning goals.
Our guiding philosophy for homework problem design was, ``{\it CM 1 is not a computational physics course. It's a course that makes use of computation.}''
Hence, most problems asked students to perform straightforward computational tasks (e.g., solve a non-stiff differential equation numerically) and to then explain the results.
We designed problems that extended analytical homework problems to non-analytic models and increased in both physical and computational sophistication as the semester progressed.
Homework problems given early in the course made explicit mention of screencasts including direct links to helpful ones. 
New computational tasks were introduced in each of the first few weeks of the course (e.g., doing algebra, plotting).
No new tasks were introduced after these first few weeks; old tasks were revisited in new physical contexts.
Later homework sets made no mention of the screencasts.


An abbreviated version of a homework problem given in the 3$^{\mathrm{rd}}$ week of the semester appears in Fig. \ref{fig:hw3}.
{Two additional homework problems appear in Appendix \ref{sec:hwapp}, and all homework problems are available online.\cite{cucomp}}
At this time, students were learning to model the drag force and to integrate the equation of motion for a single object analytically and numerically.
In this sample problem, students worked to determine if a penny dropped from the Empire State Building would kill a human being passing underneath.
This problem was one of the first where students modeled the drag force, and, hence the problem was scaffolded accordingly.
Students predicted the terminal speed (part i), made sense of the drag terms (part ii), numerically modeled the drop (part iii), compared the result (part iv) to their earlier prediction, and, finally, modeled the impact (part v).

Sandwiching computational tasks with analytical work in homework problems was a useful method for encouraging students to complete computational modeling tasks.
In prior semesters, computational modeling tasks were tacked on to the end of problems, and students would skip these tasks even though they were required.
Students completed computational tasks 25-40\% of the time in previous semesters, while 80-90\% of students now complete these tasks.

\section{Computational Final Projects}\label{sec:proj}

After a semester's worth of experience with computational modeling, students completed open-ended computational projects in the last six weeks of the course.
These projects were designed to assess students' computational fluency as well as their abilities to engage in other important scientific practices.
Moreover, we intended to motivate students to engage in authentic, scientific work.


For their projects, students selected a physical system that they found interesting, developed a computational model of this system, and wrote a report describing their model and any findings.
Most students (90\%) worked in groups consisting of no more than three students.
We provided a list of physical systems and interesting questions for our students to investigate.\cite{cucomp} 
Suggested topics included the 3-body gravitational problem (e.g., modeling stable configurations), simple routes to chaos (i.e., investigating period doubling), coupled ordinary differential equations (e.g., exploring predator/prey models), and realistic projectile motion (e.g., including spin-dependent forces).

The project was scaffolded to ensure that most students would successfully complete it.
Six weeks prior to the due date, students turned in a project description that provided details about the topic and questions they planned to investigate.
We provided feedback to students the following class day to ensure that their proposed projects were wide enough to be interesting and, yet, narrow enough that students could be reasonably expected to complete them.
Four weeks before the project was due, students prepared a short status report that described their model, provided some preliminary calculations and code, and reflected on their proposed investigations.
After reading the status reports, we met with groups who were struggling or behind schedule to develop an action plan for their project.
In the last two weeks of the course, students prepared their final reports.
We were available throughout the final two weeks to discuss projects, and a fair number of groups met with us during that period.
The vast majority of students completed the project successfully; only three students (all working alone) failed to turn in a project.

To evaluate students' reports we developed a grading rubric that scored reports and code using seven dimensions.
This rubric was made available four weeks prior to the due date. 
Students were assigned a score from 0 (poor) to 4 (excellent) on each dimension (Fig.\ \ref{fig:rubric}). 

The first four dimensions evaluated how well students mastered the course-scale learning goals (Fig.\ \ref{fig:goals}).
(1) To map the physical situation onto a computational task, a student must first demonstrate understanding of the underlying {\bf physics}.
For example, students modeling the double pendulum might describe the physical system using free-body diagrams that illustrate the important physical quantities in the problem.
(2) Typically to complete the mapping of the physical situation, a students must develop a mathematical {\bf model} of the physical system that is informed by qualitative and diagrammatic descriptions of the problem.
As an example, students who modeled the double pendulum might develop the equations of motion either from a Newtonian or Lagrangian perspective, describe the important parameters, and discuss limiting cases (e.g., small oscillations).
(3) After constructing the model, students could implement the appropriate computational method 
to produce {\bf working code}.
Students who modeled the double pendulum might use {\tt NDSolve} to numerically solve the equations of motion for a variety of initial conditions, plot the resulting motion as a function of time, and produce phase-space plots of the dynamics.
(4) After working with their code, students needed to reflect on the {\bf outcomes} of their code and to begin asking and answering interesting questions.
For example, students working with the double pendulum might begin investigating the system's sensitivity to initial conditions and discuss any interesting findings (e.g., exponential runaway of solutions).
These dimensions are closely connected to several scientific practices including developing and testing hypotheses, using and developing models, and using mathematics and computational thinking.

Student engagement in additional scientific practices (e.g., communicating sophisticated scientific information, reflecting on understanding) were evaluated by the last three dimensions.
(5) Students' reports were evaluated on aspects of the {\bf presentation} of material and ideas. (6) Students' abilities to document sources and {\bf references} were also evaluated. (7) Finally, students were evaluated on their abilities to {\bf synthesize} what they had learned by engaging in the project.
For example, students working with double pendulum might have detailed discussions about sensitivity to initial conditions, chaotic dynamics, and trajectories in phase-space.

\section{Student Performance on Final Projects}\label{sec:perf}

\begin{figure}[t]
\includegraphics[width=0.95\linewidth,clip,trim=12mm 6mm 17mm 10mm]{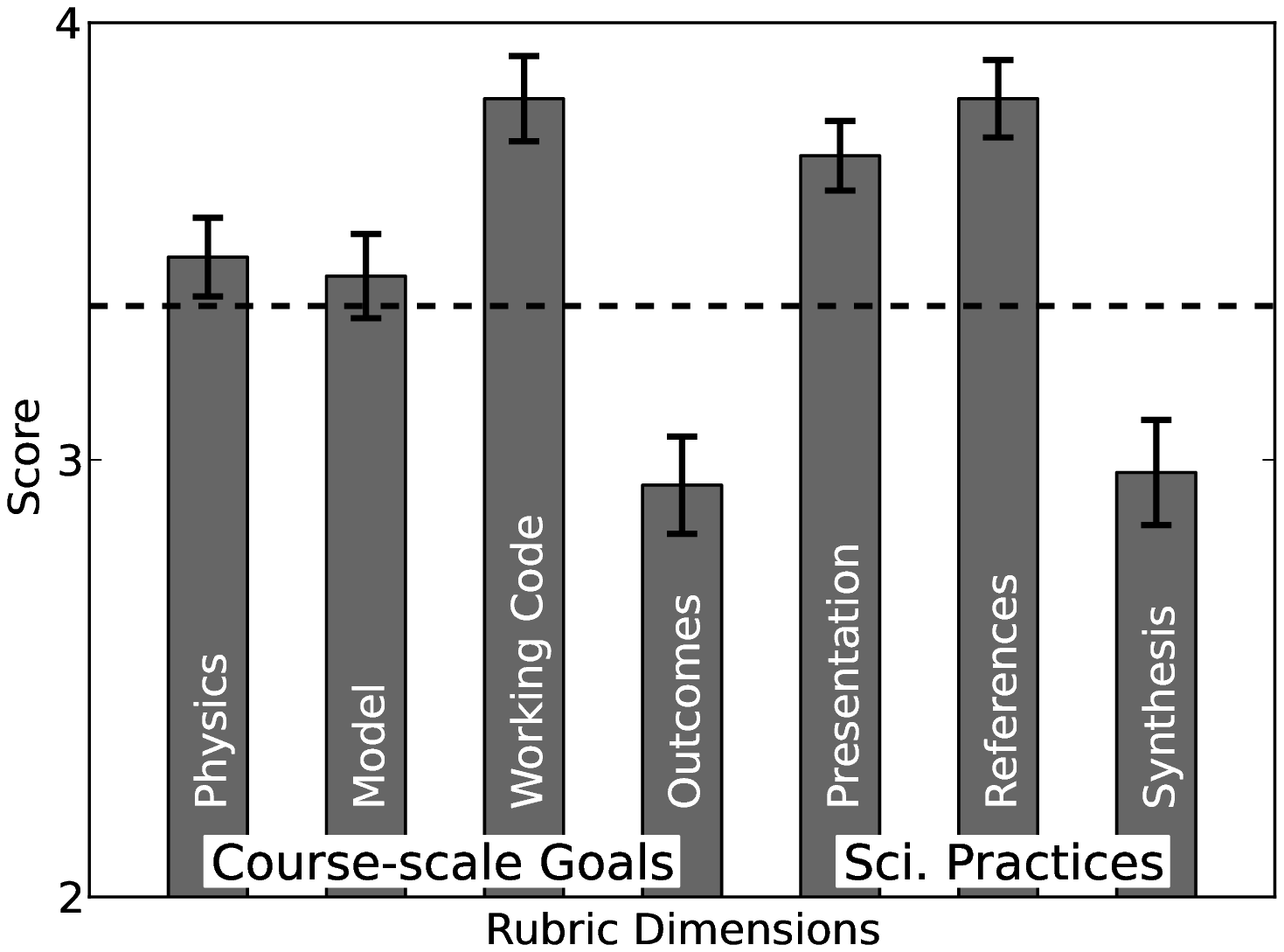}
\caption{Dimensions of the evaluation rubric on a four-point scale plotted with average scores for projects (dashed). 
}\label{fig:rubric}
\end{figure}

We were interested in how well students achieved course-scale computational learning goals and engaged in authentic, scientific practice.
Overall, student performance on the final project (avg. 83.8 $\pm$ 2.5\%) was on par with homework performance (avg. 85.9 $\pm$ 1.9\%) and exam grades (avg. 80.0 $\pm$ 1.6\%), but
what was more interesting was students' average performance on different dimensions of the rubric (Fig.\ \ref{fig:rubric}).

Students were able to successfully describe the underlying physics (Physics) and develop an appropriate mathematical model of their physical system (Model).
On their analytic course activities, students grapple with developing qualitative descriptions of the physics and connecting those descriptions to mathematical models.
Student performance suggests that such practices are transferable to the context of computational problems.

The code that students wrote to model their chosen physical system worked well and had few, if any, errors (Working Code).
However, we did not evaluate if students used elegant or efficient code nor did we give marks for discussing numerical (round-off) error because we did not emphasize such practices in instruction (Sec.\ \ref{sec:act}).
The high marks that students earned in this dimension are likely a reflection of the amount of time they devoted to writing code.
In most of our interactions, students wanted to discuss details of writing code including how to debug errors; students struggle to debug their own code when that practice is not taught explicitly.\cite{Caballero:2011tk,Caballero:2012hu}

Students found evaluating the results of their work challenging (Outcomes). 
Reflecting on the output of a computational model is a higher-level cognitive task, and such higher-level tasks are more demanding of students.\cite{Bloom:1956tx}
We suspect that students earned lower marks in this dimension because these tasks are encountered infrequently in their educational preparation.
{Similar challenges have been documented in students' analytical work in the upper-division.\cite{Caballero:2012wr,2012arXiv1207.1283W}}
Some course activities in CM 1 asked students to evaluate possible outcomes and implications, but such activities are scarce in our transformed course and are likely absent from most other courses.

Students' final reports were well-documented (References), and we were largely satisfied with students' writing (Presentation).
But, students were again challenged by high-level cognitive tasks (Synthesis).
Synthesizing information and evaluating one's own understanding of a subject are cognitively-taxing practices.
Moreover, students rarely have the opportunity to employ these practices in physics courses, even though such practices are crucial to practicing physicists.
Low marks in this dimension are representative of the lack of experience with these practices.

\section{Students' Reflections}\label{sec:refs}

{Student buy-in is critically important to sustaining any reform.\cite{dancy2013}
Documenting what students think about changes to a course is necessary to develop improvements for future iterations.}
At the end of the semester, most students (94\%) completed a survey that collected their ideas and attitudes about the course and the final project, in particular.
Students' responses to the survey suggested they were excited to explore new physics on their own, confident that they could tackle new problems, and motivated to continue in the major.

Of those who completed the survey, most students ($>$ 80\%) said the computational project gave them the opportunity to explore new physics and to use new computational techniques.
Many students chose topics that were not presented in the course because they were excited to work with new situations.
A student who investigated the Lorenz attractor noted, ``{\it the open-endedness of [the project] allowed me to investigate what I was actually interested [in], and to dabble in a field of physics that I would not normally have studied.}''
In addition, students cited that building their own computational model facilitated a deeper understanding of the physics,
``{\it I felt that an emphasis on the translation from physics on paper to physics on a computer was very important to my understanding.}''


Our model for computational instruction gave most students ($>$ 80\%) the confidence to attack open-ended problems and to explore physics on their own. 
Moreover, $\sim$90\% of our students gained confidence with Mathematica, which is important because a lack of confidence hinders student performance on computational activities.
{Computational anxiety is well-documented in computer science literature,\cite{wiebe2003computer,heersink2010measuring} and students' notions about computational modeling have been shown to affect their performance in physics.\cite{Caballero:2011tk}}
Students explicitly called out working on the final project as a confidence boost, ``{\it [the] confidence in my abilities to use Mathematica was the greatest result of this project.}"

The computational project was designed to motivate and excite students about future study in physics by engaging them in authentic, scientific practice. 
Along with lecture and homework, most students ($>$ 70\%) cited computational tasks as highly-motivating activities.
Students appreciated the opportunity to define their own project, ``{\it I liked how you allowed the students [to] decide what the project was focused on. By letting the students get excited about what they are doing, there is a better chance that they actually want to do well.}''
Engaging students in authentic, scientific practice could also have long-term effects for our students.
A female student expressed this best, ``{\it [f]or the first time in a long time, I am really really excited about physics again. I feel like I have grown so much as a physicist this semester and like I can actually succeed as an undergrad and even in graduate school/research.}''


\section{Towards a Sustainable Model for Computational Instruction}\label{sec:model}

Computational modeling is important practice for future scientists to learn; conducting modern science requires using computational tools. 
Yet, the pressure on faculty to cover course content and the lack of an effective instructional model has kept much computation from being included in students' undergraduate preparation.\cite{Chonacky:2008gq}
In this paper, we have demonstrated one model for integrating computational instruction into a typical lecture course that relies on: (1) Developing computational learning goals that reflect a basic understanding of computing (Sec.\ \ref{sec:goals}), (2) Creating instructional materials that leverage modern technologies (Sec.\ \ref{sec:act}), and (3) Assessing how students use computation in an open-ended format (Secs.\ \ref{sec:proj} \& \ref{sec:perf}).

This model has a number of affordances for non-specialist faculty and students.
Faculty who adopt this model will not need to provide computational instruction during lecture because we leveraged the ``flipped'' classroom.\cite{Lage:2000fl}
By focusing on developing students' basic computational skills, this model also does not overwhelm students, which was a major concern.
Some of our faculty teaching the second-semester classical mechanics course have built on student experience in CM 1, and designed additional computational homework problems for this follow-up course.
Our faculty value this computational experience because these activities are more closely connected to how professional physicists engage with their own work.
Moreover, such authentic work might help retain physics majors by providing enjoyable experiences (Sec.\ \ref{sec:refs}).

However, this model is not without shortcomings.
A lack of faculty expertise with Mathematica could make drawing from our limited resources challenging.
Moreover for faculty teaching other courses (e.g., electromagnetism, quantum mechanics), these resources do not yet exist.
The Partnership for Incorporating Computation in Undergraduate Physics (PICUP) is working to develop resources and activities for a broad swath of undergraduate courses using a variety of computational environments.\cite{picuplink}
But regardless of the availability of materials, core content of the course might be sacrificed when adopting this or other models.
In our experience, a computational homework question replaced a standard homework problem each week, which did not result in losing content but replacing it with something different.
However, the additional time spent on the final project caused one full week of content (i.e., separation of variables in spherical coordinates) to be dropped from the course.
We believe this was an acceptable loss given the high value that both faculty and students placed on learning computational modeling.

This model has worked well at CU Boulder, but only certain aspects of this model have been sustained.
Unfortunately, the computational final project has not been among them.
For research-focused faculty, the computational project is a significant investment of their time and energy given the large enrollment in CM 1.
Developing authentic, scientific experiences for students that can be sustained with little faculty input is challenging.
{In smaller courses or for faculty who can devote more time to instruction, adopting the final project is quite possible. Other faculty might consider using peer-review,\cite{price2012CPR} that is, having students grade each other using a modified version of our rubric. 
Alternatively, students could create short videos rather than written reports.\cite{Aiken:2013perc}

While the assessment aspect of our model still requires more testing, faculty teaching both semesters of classical mechanics have adopted the ``flipped'' model of instruction and the use of computational homework problems.
{In future iterations of CM 1, we will develop learning goals and create additional screencasts to teach motion prediction algorithms.
In addition, teaching students to evaluate their computational models will be supported by interactive course activities that emphasize basic error analysis and checking solutions.
Translating our screencasts to other popular environments (e.g., MATLAB, Python) is a longer term goal and will be completed with the help of PICUP members. 
But at our institution,} a general culture of teaching computational modeling in CM 1 and the follow-up course has taken hold.
At other institutions, local conditions will determine which aspects of this model are used, adapted, or discarded.

Developing a sustainable model of computational instruction requires others to document and share their experiences.
We must also investigate {\it how} students learn computational modeling and {\it what} students learn from computational modeling.
Enabling and understanding how students use computational modeling is largely unexplored territory for physics instruction and physics education research.
Open questions for the physics community include:
{\it (1) How do students blend conceptual physics knowledge, mathematical tools, and computational algorithms when engaging in computational modeling?} 
{\it (2) What challenges do students face when using computational modeling?} 
{\it (3) How can computational modeling be leveraged to engage students in other important scientific practices?}
Future work by the community should investigate these questions while developing and testing new models of computational instruction.
As we have seen, collaborations between traditional physics and physics education research faculty are key to supporting these changes.
Transforming undergraduate instruction is a community-driven effort and requires broad support of the community to be successful.

\begin{acknowledgments}

The authors thank J.\ Burk, J.\ Ives, A.\ Rundquist, P.\ Wagner, and B.\ Zwickl for their help drafting the initial set of computational learning goals. For providing insightful comments on an early draft of the article, we also thank S.\ Chasteen, J.\ Driscoll, K.\ Roos, and D.\ Winch. The University of Colorado's Science Education Initiative supported this work.

\end{acknowledgments}

\appendix
\section{Additional Homework Problems}\label{sec:hwapp}

Below, we provide two sample homework problems from CM 1 along with the topic-scale computational learning goals that influenced the development of each question. Each example below couples to all four course-scale learning goals. The problems below are abbreviations; full examples of all homework questions for this course are available online.\cite{cucomp}

\begin{center}
\noindent{\bf Example 1: Finding the maximum range with air resistance}
\end{center}

\noindent {\bf Computational learning goals:}\\

\noindent Plot functions of one variables.
Find the roots of an algebraic question using a root finder by setting appropriate neighborhoods.\\

\noindent {\bf Abbreviated version of homework question:}\\

\noindent We know the maximum range in vacuum for an object occurs at $\theta=\pi/4$. Estimate the maximum range (and angle) for a ball thrown with an initial speed $v_0$ in a linear medium. Consider the particular case where $v_0=v_{ter}$ and $v_{ter}^2/g=1$.\\


\noindent (a) Plot Eq.\ (2.37) in Taylor for different values of $\theta$ and determine the angle at which the range is maximum.\\
(b) Use a root finder to find the range when $\theta=\pi/4$. 
A screencast showing how to use Mathematica to find roots is available here: \url{http://youtu.be/673IQ6Z-6Yc}\\
(c) Repeat for different values of $\theta$.
Continue until you determine the maximum range and the corresponding angle to two significant figures. Compare with part (a) and the ideal case.\\
(d) What happens in the limit $v_0\gg v_{ter}$ and $v_0\ll v_{ter}$?

\begin{center}
\noindent{\bf Example 2: Investigating periods of large amplitude oscillations}
\end{center}

\noindent {\bf Computational learning goals:}\\

\noindent Integrate a complicated expression (numerically). Plot functions of one or two variables.\\

\noindent {\bf Abbreviated version of homework question:}\\

\noindent You showed that a pendulum's potential energy is $U(\phi) = mgL (1 -\cos\phi)$ and, for small angles, the motion is periodic with $\tau_0=2\pi\sqrt{L/g}$. We will now find the period for large oscillations:\\

\noindent (a) Using energy conservation, determine $\dot{\phi} (\phi)$.  Use this ODE to find an expression for the time it takes to travel from $\phi=0$ to its maximum value $\Phi_0$.  Write an expression for the full period, as a multiple of $\tau_0$.\\
(b) Evaluate the integral and plot $\tau/\tau_0$ for $0\leq \Phi_0\leq 3$ rad. For small $\Phi_0$, does your graph look like that you expect? What is $\tau/\tau_0$ for $\Phi_0=\pi/2$ rad?\\ 
(c) What happens to  $\tau$ as the amplitude of the oscillation approaches $\pi$? How well does the SHM approximate the behavior of a real pendulum?\\
(d) A real grandfather clock has a half-meter pendulum. You pull the pendulum a $\sim$10 cm to the side. Compute $\tau/\tau_0$ for this clock.   When designing the clock, would the clockmaker be safe in making the ``small angle" approximation? Why or why not?

\bibliography{comp}
\bibliographystyle{apsper}

\end{document}